\documentclass[prl,twocolumn,showpacs,preprintnumbers,amsmath,amssymb,citeautoscript]{revtex4-1}

\usepackage[dvipdfmx]{graphicx}
\usepackage{dcolumn}
\usepackage{bm}
\usepackage{color}

\newcommand{\kk}{\textbf{k}}

\newcommand{\ua}{\uparrow}
\newcommand{\da}{\downarrow}

\begin{document}

\title{Thermodynamics of transition to BCS-BEC crossover superconductivity in FeSe$_{1-x}$S$_x$}

\author{Y.~Mizukami$^1$}\email{Email: mizukami@edu.k.u-tokyo.ac.jp}
\author{M.~Haze$^2$}\thanks{Present address: Institute for Solid State Physics, University of Tokyo, Kashiwa, Chiba 277-8581, Japan} 
\author{O.~Tanaka$^1$}
\author{K.~Matsuura$^1$}
\author{D.~Sano$^2$}
\author{J.~B\"oker$^3$}
\author{I.~Eremin$^3$}
\author{S.~Kasahara$^2$}
\author{Y.~Matsuda$^2$}
\author{T.~Shibauchi$^1$}

\affiliation{
$^1$Department of Advanced Materials Science, University of Tokyo, Kashiwa, Chiba 277-8561, Japan\\
$^2$Department of Physics, Kyoto University, Sakyo-ku, Kyoto 606-8502, Japan\\
$^3$Institut fur Theoretische Physik III, Ruhr-Universitat Bochum, D-44801 Bochum, Germany\\
}

\date{\today}

\begin{abstract}

The BCS-BEC crossover from strongly overlapping Cooper pairs to non-overlapping composite bosons in the strong coupling limit has been a long-standing issue of interacting many-body fermion systems. Recently, FeSe semimetal with hole and electron bands emerged as a high-$T_{\rm c}$ superconductor located in the BCS-BEC crossover regime, owing to its very small Fermi energies. In FeSe, however, an ordinary BCS-like heat-capacity jump is observed at $T_{\rm c}$, posing a fundamental question on the characteristics of the BCS-BEC crossover. Here we report on high-resolution heat capacity, magnetic torque, and scanning tunneling spectroscopy measurements in FeSe$_{1-x}$S$_x$. Upon entering the tetragonal phase at $x>0.17$, where nematic order is suppressed, $T_{\rm c}$ discontinuously decreases. In this phase, highly non-mean-field behaviors consistent with BEC-like pairing are found in the thermodynamic quantities with giant superconducting fluctuations extending far above $T_{\rm c}$, implying the change of pairing nature. Moreover, the pseudogap formation, which is expected in BCS-BEC crossover of single-band superconductors, is not observed in the tunneling spectra. These results illuminate highly unusual features of the superconducting states in the crossover regime with multiband electronic structure and competing electronic instabilities.

\end{abstract}



\maketitle

\section{I. INTRODUCTION}
\begin{figure*}[t]
\centering
\includegraphics[width=0.8\linewidth]{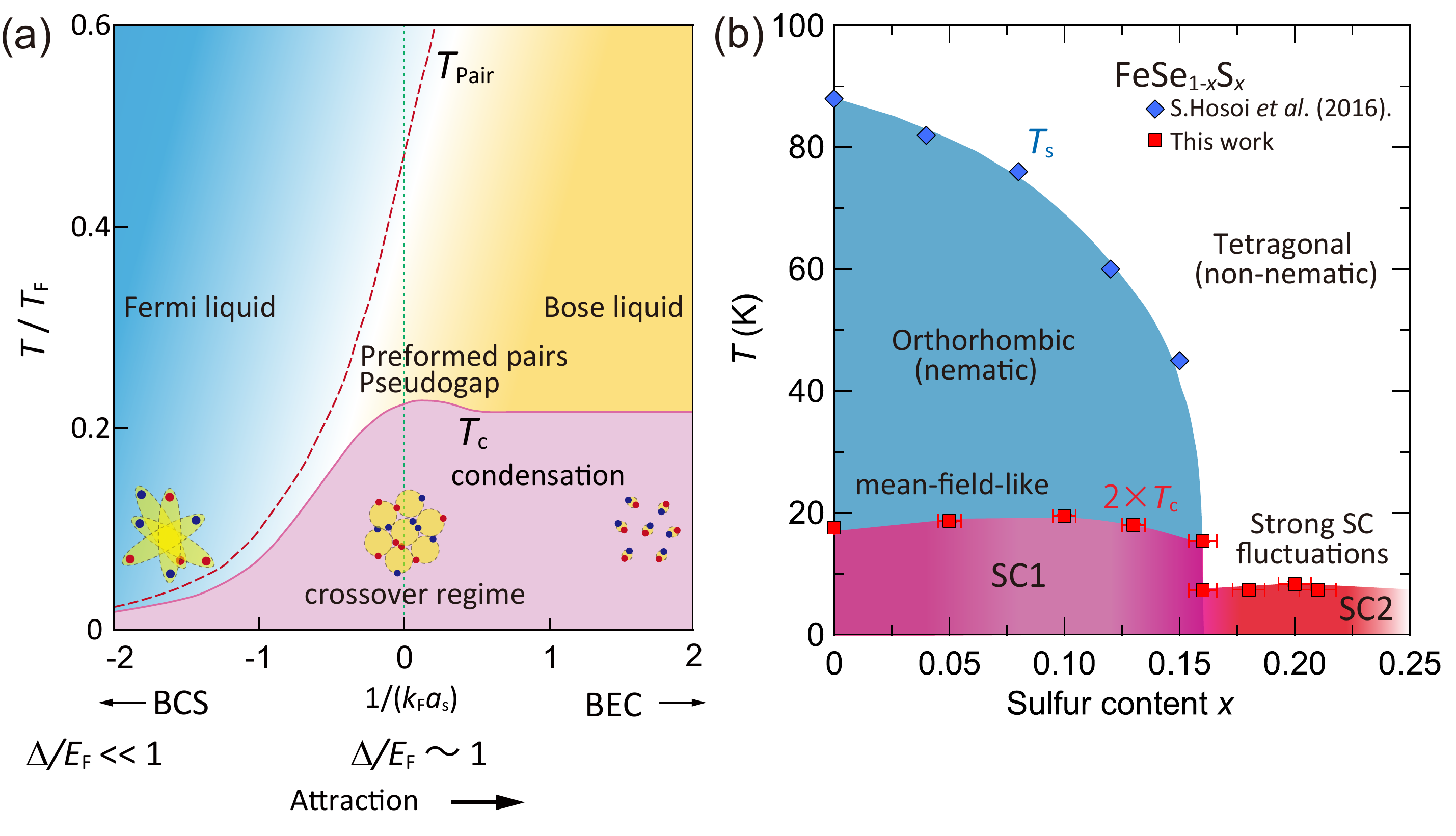}
\vspace{1.5mm}
\caption{Phase diagrams of BCS-BEC crossover in single-band systems and multiband FeSe$_{1-x}$S$_x$ superconductors.
	(a) Theoretical phase diagram of BCS-BEC crossover generally obtained for single-band systems~\cite{melo08,randeria14}. The temperature $T$ is normalized by the Fermi temperature $T_{\rm F}$ and the strength of attraction is given by dimensionless coupling constant $1/(k_{\rm F}a_s)$. Below the pairing temperature $T_{\rm pair}$, bosonic pairs form, while the superconducting coherence is acquired when the condensation occurs at $T_{\rm c}$. In the weak-coupling BCS limit, $\Delta/\varepsilon_{\rm F}$ is much smaller than unity, and the coherence length $\xi$ is much longer than the interparticle distance ($\sim 1/k_{\rm F}$) and $T_{\rm pair}$ is very close to the actual $T_{\rm c}$. In the BCS-BEC crossover regime $\Delta/\varepsilon_{\rm F}\sim 1$, $T_{\rm c}/T_{\rm F}$ reaches a maximum, and preformed Cooper pairs are expected to exist in an extended temperature region between $T_{\rm pair}$ and $T_{\rm c}$. 
	(b) Experimentally determined $T$-$x$ phase diagram of FeSe$_{1-x}$S$_x$. The superconducting transition temperature $T_{\rm c}$ (red squares) is determined by the present heat capacity measurements using small crystals. The nematic transition temperature (blue diamonds) determined by transport measurements~\cite{hosoi16} is also plotted. The abrupt change in $T_{\rm c}$ indicates a significantly different superconducting ground states (SC1 and SC2) divided by the nematic end point at $x_{\rm c}\approx 0.17$.
} 
\label{fig1}
\end{figure*}

The crossover between weak-coupling BCS theory of superconductivity and strong-coupling BEC describes the fundamentals of quantum bound states of particles, including information on the superconducting critical temperature ($T_c$) and the possible pseudogap formation~\cite{melo08,randeria14}. In the strong-coupling regime, the pair-formation temperature $T_{\rm pair}$ and the actual superconducting transition temperature $T_{\rm c}$ are distinctly separated (Fig.\:\ref{fig1}(a)). Between $T_{\rm pair}$ and $T_{\rm c}$, the so-called preformed Cooper pairs exist, which give rise to large superconducting fluctuations and possibly a depletion of the low-energy density of states (DOS), namely the pseudogap. Almost all experimental studies on the crossover physics have been performed in ultracold atomic systems for the past decades~\cite{randeria14}, because of the difficulty in tuning the attractive interaction between electrons in solids. Therefore, it is extremely challenging to realize such strongly-coupled pairs in electron systems. In underdoped high-$T_{\rm c}$ cuprates, the pseudogap formation has been discussed in terms of the BCS-BEC crossover~\cite{chen}, but recent experiments support a phase transition of different kinds at the onset of pseudogap and the relevance of BCS-BEC crossover in cuprates remains unresolved~\cite{Keimer}.

Recently the superconducting semimetal FeSe~\cite{hsu08,boehmer13,shibauchi20} with very small hole and electron Fermi surfaces is found to exhibit large ratios of the superconducting energy gap $\Delta$ and Fermi energy $\varepsilon_{\rm F}$, $\Delta/\varepsilon_{\rm F} \approx 0.3$-1.0~\cite{kasahara14,terashima14,rinott17}, which appears to place FeSe in the crossover regime (Fig.\:\ref{fig1}(a)). Indeed, in high-quality single crystals with long mean free paths of electrons~\cite{kasahara14}, large superconducting fluctuations have been reported from the torque magnetometry, providing support for the presence of fluctuated Cooper pairs created above $T_{\rm c}$ in the crossover regime~\cite{kasahara16}. These observations underline the importance of the BCS-BEC crossover physics both in the superconducting and normal states of FeSe.
On the other hand, scanning tunneling spectroscopy (STS) measurements do not observe the pseudogap formation above $T_{\rm c}$~\cite{hanaguri19}.  

These results imply that the superconductivity in FeSe is not simply described by a  picture of the conventional BCS-BEC crossover. An important aspect which has not been taken into account in the previous studies is the effects of multiband electronic structure, which may have other tuning factors of the crossover in addition to $\Delta/\varepsilon_{\rm F}$. In fact, Fermi surface of  FeSe is composed of separated hole and electron pockets with different values of $\Delta/\varepsilon_{\rm F}$~\cite{kasahara14,terashima14,coldea17}. It has been pointed out that in the multiband systems the developments of superconducting fluctuations and pseudogap formation are sensitively affected by the interband coupling strength~\cite{salasnich18}, indicating that the normal and superconducting properties associated with the crossover in multiband systems can be different from those in the single band systems~\cite{chubukov16}. It is quite important, therefore, to clarify how the superconducting properties evolve when the multiband electronic structure is tuned.

Here, we focus on FeSe$_{1-x}$S$_x$, where the orthorhombic (nematic) phase transition at 90\,K in FeSe can be tuned by isovalent sulphur (S) substitution. The nematic transition that distorts the Fermi surface is completely suppressed at a critical concentration $x_{\rm c} \approx 0.17$~\cite{hosoi16,Licciardello19}. The superconductivity persists in the tetragonal (non-nematic) phase above $x_{\rm c}$.
In the case of applying hydrostatic pressure, the suppression of nematic order is accompanied by the pressure-induced antiferromagnetic order~\cite{matsuura17}. In contrast, S substitution suppresses the nematic order without inducing the antiferromagnetism that significantly changes the Fermi surface through band folding even when crossing $x_c$. This enables us to investigate the evolution of superconducting state in the BCS-BEC crossover with little influence of antiferromagnetic fluctuations. Moreover, the observations of quantum oscillations up to $x \approx 0.19$~\cite{coldea19} demonstrate that the S ions do not act as strong scattering centers.

\begin{figure*}[t]
\centering
\includegraphics[width=1.0\linewidth]{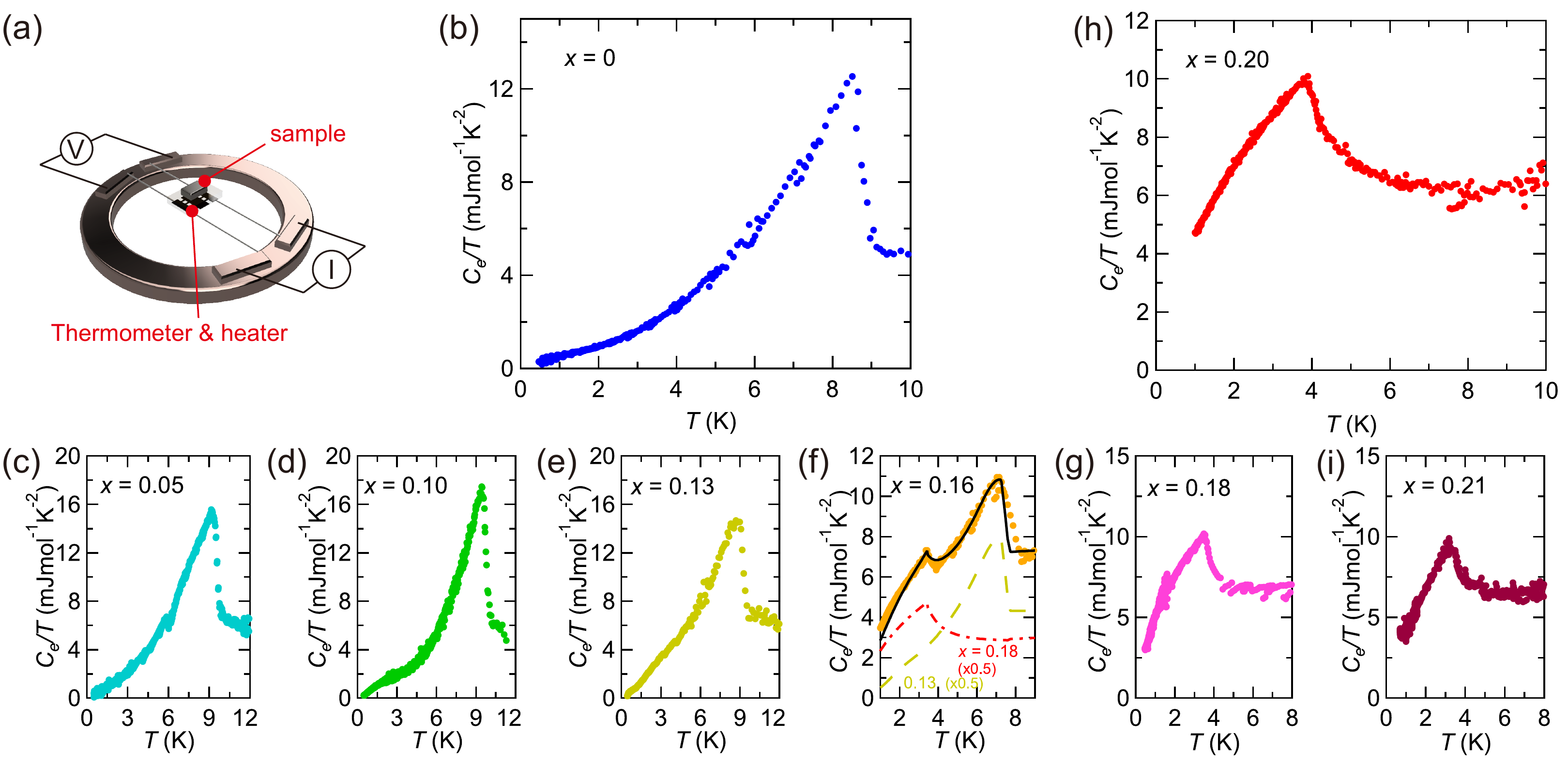}
\vspace{-3mm}
\caption{Heat capacity in FeSe$_{1-x}$S$_x$ superconductors.
	(a) The schematic of experimental set up. A Cernox small thermometer chip is used as a heater and large current pulses are applied.
      (b)-(i) $T$-dependence of electronic heat capacity divided by $T$, $C_{\rm e}/T$ in FeSe$_{1-x}$S$_x$ single crystals at low temperatures for $x= 0$ (b), $x= 0.05$ (c), $x= 0.10$ (d), $x= 0.13$ (e) in the orthorhombic phase and $x= 0.18$ (g), $x= 0.20$ (h), $x= 0.21$ (i) in the tetragonal phase.
	(f) $T$-dependence of $C_{\rm e}/T$ for $x=0.16$ near the nematic end point. The data (orange dots) are fitted with a weighted sum (black line) of the $x= 0.13$ orthorhombic data (dashed line) and $x= 0.18$ tetragonal data (dashed-dotted line) with adjusted $T_{\rm c}$ values.   
}
\label{fig2}
\end{figure*}

\section{II. HEAT CAPACITY}

Through the high-resolution heat capacity, magnetic torque, and STS measurements on high-quality single crystals, we investigate the evolution of superconducting and normal state properties with $x$ in FeSe$_{1-x}$S$_x$. In Fig.\:\ref{fig1}(b), we show the phase diagram obtained by the heat capacity measurements. The drastic change in $T_c$ and superconducting fluctuations are found across $x_c$. To ensure the sample homogeneity, we used very small crystals (10-190 $\mu$g) for the thermodynamic measurements, and employed the long relaxation method with a home-made cell (Fig.\:\ref{fig2}(a)) designed for small single crystals. Figures\:\ref{fig2}(b)-(e) depict the $T$-dependence of electronic heat capacity divided by $T$, $C_{\rm e}/T$, for $x = 0$, 0.05, 0.10 and 0.13. Here we subtract the phonon contribution, which shows $T^3$-dependence, as confirmed by the measurements in the normal state at high fields (Supplementary Materials). In Fig.\:\ref{fig2}(b), $C_{\rm e}/T(T)$ in FeSe ($x=0$) exhibits an ordinary BCS-like jump at $T_{\rm c}=9$\,K, consistent with the previous studies on larger samples~\cite{lin11,yang17,hardy19}. As the temperature is lowered below $T_{\rm c}$, $C_{\rm e}/T$ decreases with decreasing slope. Below $\sim 3$\,K, $C_{\rm e}/T$ shows $T$-linear dependence, which is consistent with the strongly anisotropic superconducting gap~\cite{kasahara14,hardy19,sprau17}. Similar $C_{\rm e}/T(T)$ is observed in orthorhombic phase $x=$0.05, 0.10, and 0.13 with $T_{\rm c}$ of 9.7, 10 and 9.5\,K, respectively (Figs.\:\ref{fig2}(c), (d) and (e)).

Upon entering the tetragonal phase at $x>x_{\rm c}$, $C_{\rm e}/T$ exhibits a dramatic change not only in the superconducting state but also in the normal state. As represented by $x$ = 0.20 in Fig.\:\ref{fig2}(h), $C_{\rm e}/T$ increases with increasing slope in the normal state when approaching $T_{\rm c}$ from high temperatures.  In addition,  $C_{\rm e}/T$ exhibits a kink at $T_{\rm c}$,  without showing a discontinuous jump. Such a continuous $T$-dependence at $T_{\rm c}$ is reminiscent of the BEC transition in the free Bose gas systems~\cite{pethick08} although it is too simple to employ the free Bose gas model to our system. We note that similar enhancement of $C_{\rm e}/T$ in the normal state is obtained in the recent calculations for the strongly interacting Fermi systems near the unitary limit in the BCS-BEC crossover regime~\cite{Wyk16}. As the temperature is lowered below $T_{\rm c}$, $C_{\rm e}/T(T)$ for $x = 0.20$ decreases with increasing slope, in stark contrast to $C_{\rm e}/T$($T$) in the orthorhombic crystals as shown in Figs.\:\ref{fig2}(b)-(e). Below $T_{\rm c}$, $C_{\rm e}/T$ decreases nearly in proportion to $T^{1/2}$ with large residual $C_{\rm e}/T$ in the limit $T \rightarrow 0$, which is indicative for a large DOS at zero energy. The large residual DOS is consistent with the recent thermal conductivity and STS measurements~\cite{sato18, hanaguri18}. Similar $C_{\rm e}/T(T)$ is also observed for $x$ = 0.18, and 0.21 in the tetragonal phase as shown in Figs.\:\ref{fig2}(g), and (i). These results together with the sharp change in $T_c$ across $x_{\rm c}$ demonstrate that the superconducting properties in the tetragonal phase are drastically different from those in the orthorhombic phase.

We stress that the observed peculiar $C_{\rm e}/T(T)$ in tetragonal phase does not stem from chemical inhomogeneity within the small sample for the following reasons. First, topographic image in STM of FeSe$_{1-x}$S$_x$ crystals reveal no evidence for the segregation of S atoms, indicating an excellent homogeneity~\cite{hanaguri18} (see also Figs.\:\ref{fig3}(a) and (b)). Second, the observation of the quantum oscillations of the crystals in the same batch~\cite{coldea19} demonstrates a long mean free path of the carriers, ensuring the high quality of our crystals. Third, according to the elemental mapping of the energy dispersive X-ray spectroscopy (EDX) at different scales of distance (see Fig.S1), the typical spatial variation of the composition $\Delta x$ is $\sim$~0.01, and there are no discernible segregations or large inhomogeneity of the chemical compositions which can explain our observation in $C_{\rm e}/T(T)$. Fourth, we measured the heat capacity on two different crystals for $x=0.20$(see Fig.\,S2), and two other concentrations $x=0.18$ and 0.21(Figs.\:\ref{fig2}(g), and (i)), and confirmed the reproducibility of the behavior. We also note that our results are consistent with the previous data of the crystal with much larger volume within the error bar (see Fig.\,S2). Fifth, most importantly, there is no concentration of the sample, which shows the superconducting transition between $\approx 4$\,K and $\approx 7$\,K in the phase diagram of FeSe$_{1-x}$S$_x$. This is supported by $C_{\rm e}/T(T)$ of the crystal at $x\approx 0.16$, which is located in the very vicinity of $x_{\rm c}$.  As shown in Fig.\:\ref{fig2}(f), $C_{\rm e}/T$ at $x\approx 0.16$ exhibits two well separated peaks at $T_{\rm c1}=4$\,K and  $T_{\rm c2}=7$\,K, indicating a phase separation (due to the small spatial variation of $\Delta x\sim0.01$) at $x_c$. This $T$-dependence is well reproduced by the sum of $C_{\rm e}/T$ of $x=0.13$ ($T_{\rm c}=9.5$\,K) and 0.18 ($T_{\rm c}=4.0$\,K), assuming that both contribute equally. Here, for the fitting, $T_{\rm c}$ values are shifted slightly. These results indicate a discontinuous change of $T_{\rm c}$ at $x_{\rm c}$. Such a discontinuous change in the superconducting properties is consistent with the jump in $\Delta$ values at $x_{\rm c}$ recently reported by STS studies~\cite{hanaguri18}. The observed enhancement of $C_{\rm e}/T$ above $T_{\rm c}$ for the sample in tetragonal phase, therefore, can be attributed to an intrinsic electronic property. 

\section{III. SCANNING TUNNELING MICROSCOPY/SPECTROSCOPY}

\begin{figure}[t]
\centering
\includegraphics[width=0.9\linewidth]{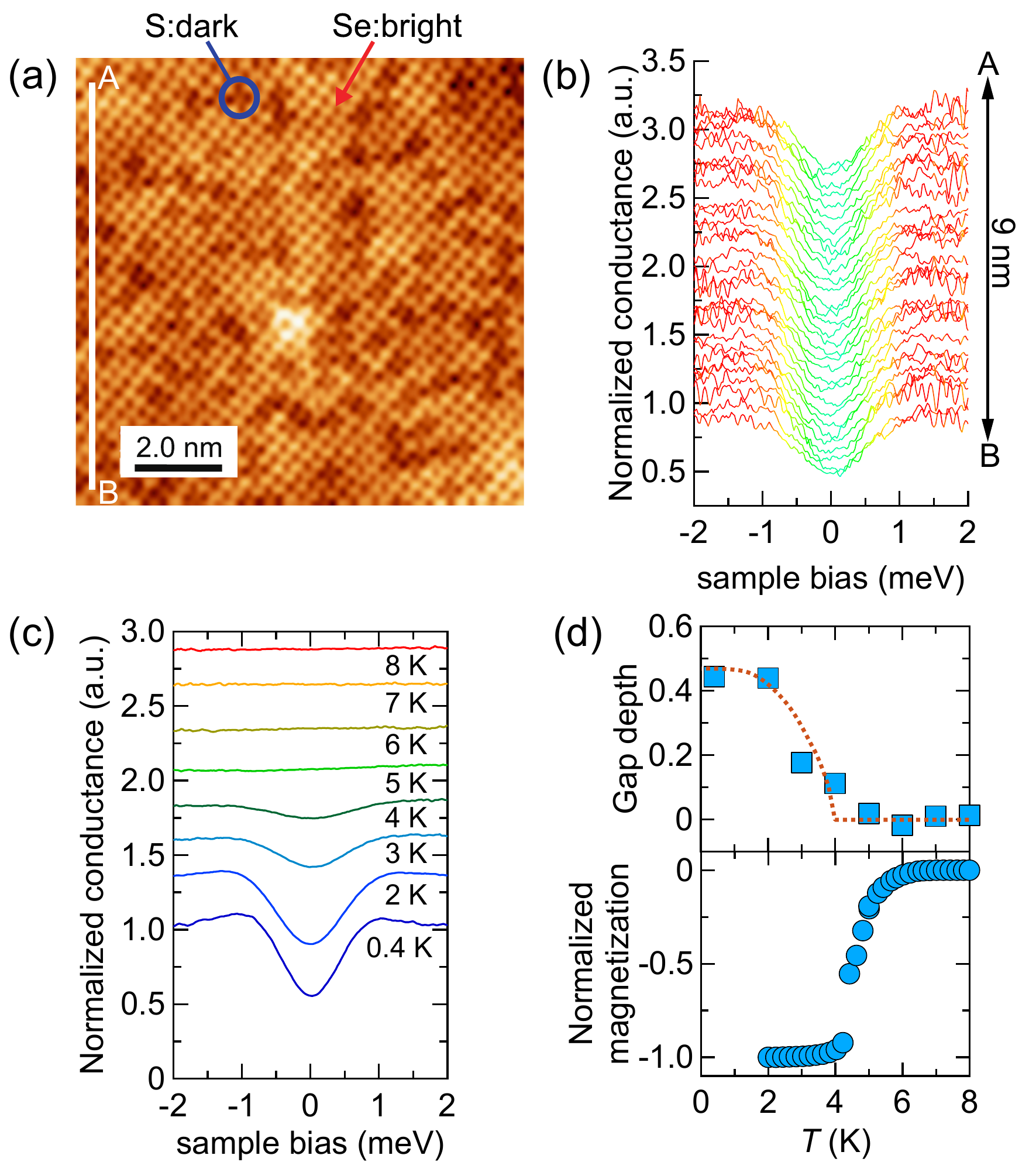}
\vspace{-3mm}
\caption{Scanning tunneling spectroscopy in the tetragonal phase.
	(a) Topographic STM image (10~mV, 100~pA) of FeSe$_{0.75}$S$_{0.25}$ surface over a 10~nm by 10~nm area.
     (b) A line profile of the normalized conductance along the AB shown in {\bf a}. The data are taken at 2~K.
	(c) Normalized tunneling conductance spectra taken at several temperatures. The spectra are vertically shifted for clarity. The tip is stabilized at $V$ = 6~mV and $I$ = 1~nA.
	(d) Upper panel shows the $T$-dependence of gap depth, which is defined by the difference between normalized conductance values at $V$ = 0 and 6~mV. The red dashed line is a guide for the eye. Lower panel shows the $T$-dependence of magnetization measured at a zero field cool condition.
}
\label{fig3}
\end{figure}

To elucidate whether the depletion of DOS associated with the pseudogap formation occurs,  the STS measurements on the crystal of $x=0.25$ are performed. By counting the number of S atoms in the STM topographic image, we determined $x$ value accurately (Fig.\:\ref{fig3}(a)). Figure\,\ref{fig3}(b) shows the line profile of the normalized conductance along the line AB depicted in Fig.\:\ref{fig3}(a). Spectra show little variation within 10 nm scale, demonstrating the uniform spatial distribution of the superconducting gap. We also note that the gap value is consistent with the previous report~\cite{hanaguri18}. Figure\,\ref{fig3}(c) depicts the STS spectra normalized by the conductance above the superconducting gap in a wide temperature range. Large residual DOS, which is more than half of the normal state value outside the superconducting gap ($\Delta \approx 1$\,{meV}) is observed even at $T = 0.4$\,K ($T/T_{\rm c} = 0.1$). This large residual DOS is consistent with the large $C_{\rm e}/T$ at low temperatures. Remarkably, STS spectrum becomes nearly energy independent at high temperatures above $\sim 5$\,K, showing no signature of the gap formation. This is also seen by the $T$-dependence of the gap depth plotted in the upper panel of Fig.\:\ref{fig3}(d). The superconducting gap closes at $\approx 5$\,K below which the $T$-dependence of magnetization shows a rapid decrease as shown in the lower panel of Fig.\:\ref{fig3}(d).


\begin{figure*}[t]
\centering
\includegraphics[width=0.9\linewidth]{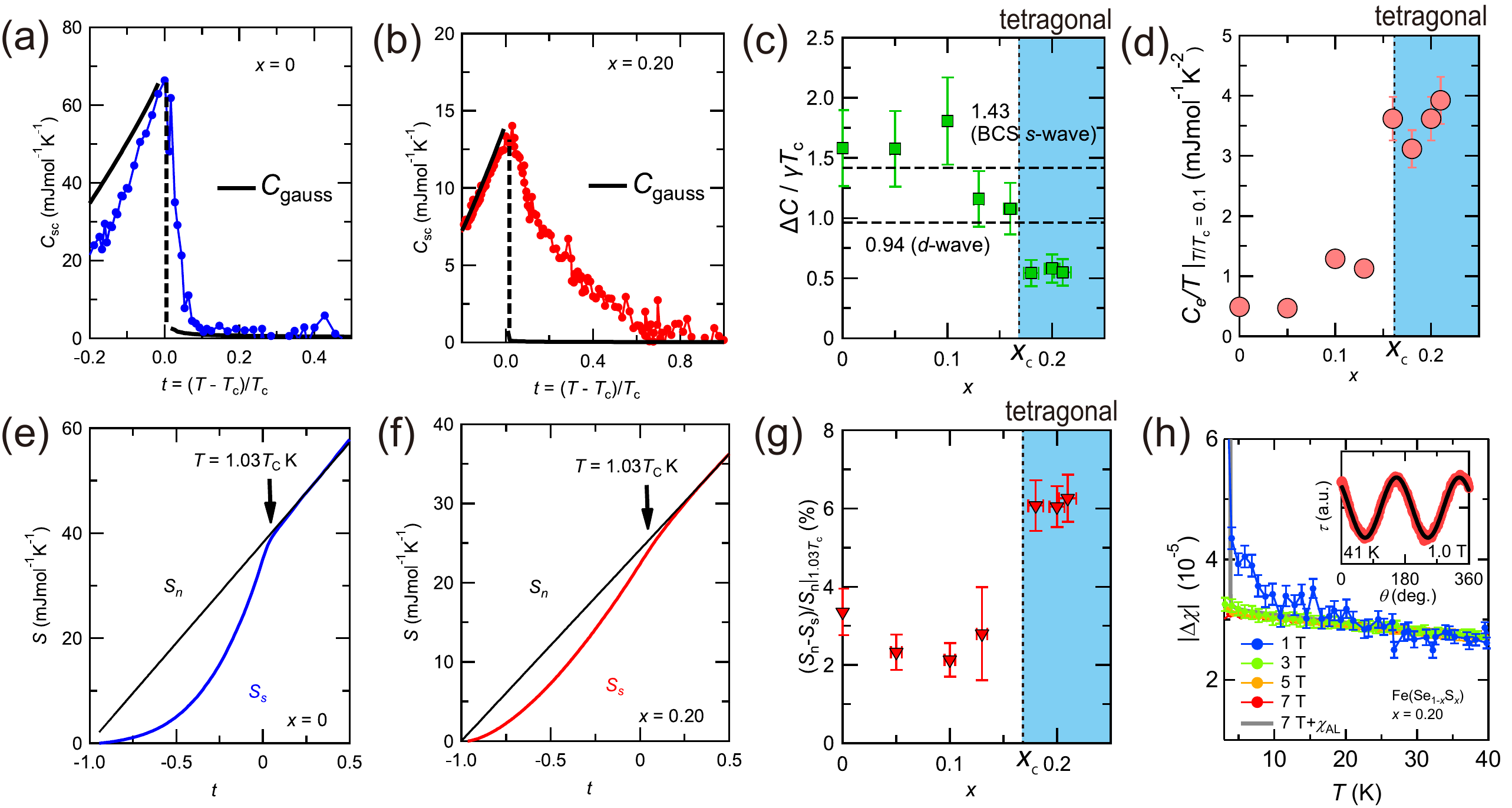}
\vspace{-3mm}
\caption{Comparisons of superconducting transition properties between the orthorhombic and tetragonal phases.
	(a) The superconducting contribution to the electronic heat capacity $C_{\rm sc}=C_{\rm e}-\gamma T$ as a function of reduced temperature $t \equiv \frac{T-T_{\rm c}}{T_{\rm c}}$ for $x=0$ ({\bf ($A$)}) and $x=0.20$ ({\bf ($B$)}). The conventional Gaussian fluctuation contribution $C_{\rm gauss}$ is estimated (solid lines).
	(c), (d) $x$ dependence of $\Delta C/\gamma T_{\rm c}$ ({\bf ($C$)}), and $C_e/T$ at $T/T_c$ = 0.1 ({\bf ($D$)}).
	(e), (f) Entropy calculated from $C_{\rm e}/T(T)$ in the normal state ($S_{\rm n}$) and superconducting state ($S_{\rm n}$) as a function of $t$ for $x=0$ ({\bf ($E$)}) and $x=0.20$ ({\bf ($F$)}). $S_{\rm n}$ is estimated from the high-field data where superconductivity is suppressed. The entropy for $x=0.20$ is linearly extrapolated below 1.0~K.
	(g) $x$ dependence of the relative entropy difference between zero and high fields $(S_{\rm n}-S_{\rm n})/S_{\rm n}$ at $T=1.03T_{\rm c}$.
	(h) The $T$-dependence of the anisotropy of the magnetic susceptibility $\Delta\chi = \chi_c - \chi_{ab}$ for several fields in $x$ = 0.20. Each data is obtained by fitting the field-angle dependence of the torque $\tau$($\theta$). The inset shows the $\tau$($\theta$) at 1.0~T and 41~K as the typical signal of torque, where the red markers and the solid black line show the raw data and fitting curve, respectively.
}
\label{fig4}
\end{figure*}

Nearly energy-independent DOS at $\varepsilon_{\rm F}$ observed by STS above $T_{\rm c}$ indicates that the observed enhancement of $C_{\rm e}/T$ above $T_{\rm c}$ in tetragonal phase is not caused by the DOS effect. Therefore, it is natural to consider that the enhancement of $C_{\rm e}/T$ stems from the superconducting fluctuations. Indeed, similar enhancement of $C_{\rm e}/T$ toward $T_c$ due to the strong fluctuations has been previously reported in other iron-based superconductors~\cite{pribulova09, serafin10, welp11}. In order to discuss the effect of the superconducting fluctuations in FeSe$_{1-x}$S$_x$ quantitatively,  we extract the superconducting contribution $C_{\rm sc}$, which is obtained by subtracting the normal state electronic contribution $\gamma T$ from $C_{\rm e}$ for $x$=0 (blue circles in Fig.\:\ref{fig4}(a)) and $x$=0.20 (red circles in Fig.\:\ref{fig4}(b)). We compare $C_{\rm sc}$ with the conventional term of the mean field Gaussian fluctuations~\cite{inderhees88}(Supplementary Materials). Obviously, the heat capacity contribution that significantly exceeds $C_{\rm gauss}$ can be seen for $x=0.20$. This extra heat capacity is observable up to $t\sim 0.8$. In Fig.\:\ref{fig4}(c), we display the $x$-dependence of $\Delta C/\gamma T_{\rm c}$, where $\Delta C$ is the height of $C_{\rm sc}$ at $T_{\rm c}$. For $x<x_{\rm c}$, the magnitude of $\Delta C/\gamma T_{\rm c}$ is close to the BCS weak coupling value (1.43 for $s$-wave and 0.94 for $d$-wave). On the other hand, $\Delta C/\gamma T_{\rm c}$ for $x>x_{\rm c}$ is reduced far below the mean-field BCS value. As a consequence of unusual suppression of $\Delta C/\gamma T_c$, the low-$T$ $C_{\rm e}/T$ is largely enhanced through the entropy balance. Figure\:\ref{fig4}(d) depicts the $x$-dependence of the $C_e/T$ at $T/T_c$ = 0.1 taken from Fig.\:\ref{fig2}. Remarkable enhancement of $C_e/T$ at $x_c$ indicates the fundamental difference of the low-energy excitations across $x_c$. We note that recent NMR measurements for $x>x_c$ report no splitting of the spin-echo signal in the crystal from the same batch~\cite{wiecki18,kuwayama20}, which excludes the microscopic phase separation of superconducting and non-superconducting regions.

\section{IV. MAGNETIC TORQUE}

The anomalous $C_{\rm e}/T$ in the tetragonal phase is reflected by the entropies in the normal and superconducting states, $S_{\rm n}$ and $S_{\rm s}$, respectively, as shown in Fig.\:\ref{fig4}(e) and (f). For $x = 0$, the $T$-dependence of $S_{\rm s}$ shows a kink at $T_{\rm c}$, which is typical for the second-order superconducting transition. For $x = 0.20$, on the other hand, owing to the absence of the jump in $C_{\rm e}/T$ at $T_{\rm c}$, no kink anomaly is observed in the $T$-dependence of $S_s$. Moreover, the entropy is significantly suppressed from $S_n$ even above $T_{\rm c}$. In Fig.\:\ref{fig4}(f), we show  the $x$ dependence of the lost entropy normalized by $S_{\rm n}$, $(S_{\rm n}-S_{\rm s})/S_{\rm n}$, just above $T_{\rm c}$, $T=1.03T_{\rm c}$. Upon entering the tetragonal phase, an abrupt enhancement of $(S_{\rm n}-S_{\rm s})/S_{\rm n}$ can be seen, indicating a drastic change of the superconducting fluctuations associated with the preformed Cooper pairs. 

To obtain further insight into the superconducting fluctuations in the tetragonal phase, the magnetic torque was measured by using micro-cantilevers. The torque can sensitively detect the diamagnetic response through the anisotropy of magnetic susceptibility~\cite{kasahara16}.  By measuring the field angular variation of torque as a function of polar angle $\theta$ from the $c$ axis,  the anisotropy $\Delta \chi=\chi_c-\chi_{ab}$ is determined, where $\chi_c$ and $\chi_{ab}$ are the magnetic susceptibilities along the $c$ axis and in the $ab$ plane, respectively. As shown in the inset of Fig.\:\ref{fig4}(h), $\tau(\theta)$ shows an almost perfect sinusoidal curve, $\tau(\theta)\propto\Delta\chi \sin \theta$. Although $\Delta\chi(T)$ at high fields is almost independent of $T$ and $H$, it exhibits strong $T$ and $H$-dependence at low fields as shown in Fig.\:\ref{fig4}(h). It is well settled that the fluctuation induced diamagnetic susceptibility of most superconductors including multiband systems can be well described by the standard Gaussian-type (Aslamasoz-Larkin, AL) fluctuation susceptibility $\chi_{\rm AL}$(Supplementary Materials). Here we focus on the $H$-dependent diamagnetic susceptibility at low fields\cite{kasahara16}. Then, $\Delta\chi$ at low fields is much larger than the $|\chi_{\rm AL}|$ which is added to high field 7~T data. This implies that the superconducting fluctuations in FeSe$_{1-x}$S$_x$ are distinctly different from those in conventional superconductors, supporting the fluctuation effects observed in heat capacity measurements. We stress that although the multi-gap superconductivity may lead to a small value of $\Delta C/\gamma T$, it cannot give rise to the non-mean-field behaviors observed in heat capacity and magnetic susceptibility above $T_c$.


\section{V. DISCUSSION AND CONCLUSION}

In FeSe, while large superconducting fluctuations that well exceed the Gaussian fluctuations are observed, the $C_{\rm e}/T$ around $T_c$ does not exhibit a significant deviation from the mean-field behavior. This suggests that tetragonal FeSe$_{1-x}$S$_x$ are closer to the BEC regime. However, with increasing S composition, the volume of the Fermi surface increases, as reported by quantum oscillation experiments~\cite{coldea19}, and $\Delta$ is reduced, according to STM measurements~\cite{hanaguri18}. Then $\Delta/\varepsilon_F$ is expected to become smaller with increasing S concentration, which is opposite to the tendency of approaching the BEC regime~\cite{hashimoto20}. This suggests that there is another important factor, which has not been duly taken into consideration.

A key factor that controls the BCS-BEC crossover in FeSe$_{1-x}$S$_x$ is the multiband character. As shown previously~\cite{chubukov16,hanaguri19}, in FeSe the strong interband interaction between electron and hole pockets prevents the splitting of the $T_{\rm pair}$ and the $T_c$, determined by the superfluid stiffness. We note that this strong interband interaction is also an important ingredient in the orbital-selective scenario~\cite{sprau17,yu18,hu18, nica21} and therefore we expect that even taking orbital selectivity into account, the main result, {\it i.e.} mean-field like jump of the specific heat will remain the same. At the same time, much less is known on the evolution of the ratio of the interband versus intraband interactions upon approaching $x_c$ in FeSe$_{1-x}$S$_x$. Furthermore, recent experimental~\cite{Yi2019} and theoretical~\cite{Rhodes2020,Steffensen2020} works pointed out an interesting possibility of the substantial non-local nematic order on the $d_{xy}$ orbital and/or inter-orbital-nematicity between $d_{xy}$ and $d_{yz}$ orbitals in FeSe. The most important consequence of this phenomenon would be an additional Lifshitz transition in FeSe$_{1-x}$S$_x$ upon S substitution, which occurs close to $x_c$. This in turn implies the appearance of the incipient band near $x_c$ (see for example Fig.6 in Ref.\cite{Rhodes2020}). On general grounds, the presence of the incipient band would enhance the tendency towards the BCS-BEC crossover. In addition, given nearly equal mixture of the $s$- and $d$-wave symmetry components of the superconducting gap in FeSe~\cite{hashimoto18} to express such anisotropic gap, it is natural to assume that they are competing in the tetragonal phase of FeSe$_{1-x}$S$_x$. In particular, the ratio of the pairing interactions in the $d$-wave and $s$-wave channels $|g^d_{\rm ee}|/g^s_{\rm eh}$ may, in addition to the incipient band, be another key parameter of the BCS-BEC crossover as follows from our model calculations (see Fig.\:S4). The pair-formation temperature $T_{\rm pair}$ and condensation temperature $T_{\rm c}$ can be split more with increasing the ratio of intraband and interband interactions $|g^d_{\rm ee}|/g^s_{\rm eh}$ (see Fig.\:S4(b)) without changing $\varepsilon_{\rm F}$. This quantitatively explains the appearance of strong coupling superconductivity for $x>x_c$ with no apparent enhancement of $\Delta/\varepsilon_F$. A remarkable and unexpected feature is the absence of pseudogap above $T_{\rm c}$ in the STS spectra, despite the presence of giant superconducting fluctuations in thermodynamic quantities. It should be noted that the relationship between the preformed pairs and the pseudogap formation is still elusive even in the ultra-cold atoms~\cite{randeria14}. In addition, the onset temperature of the pseudogap may differ from that of fluctuations~\cite{Wyk16}. Moreover, it has been pointed out that the pseudogap phenomena in multiband systems are markedly altered from those in the single band systems. For instance, in multiband systems, the pseudogap formation is less prominent by the reduction of its onset temperature~\cite{tajima20}. Quite recently, the BEC-like dispersion of the Bogoliubov quasiparticles has been reported in the tetragonal FeSe$_{1-x}$S$_x$ by using ARPES measurements~\cite{hashimoto20}. Although the direct comparison of the ARPES data with the present thermodynamic results is not straightforward, because ARPES measures only the portion of hole Fermi pocket, it provides spectroscopic evidence of the BCS-BEC crossover. The ARPES measurements also report the presence of distinct pseudogap in the hole pocket. However, we point out that this is not an apparent contradiction to the present STM results. It has been suggested that the DOS of electron pocket is much larger than that of hole pockets. As STM measures the DOS integrated over all bands, STM is insensitive to the pseudogap formation in the hole pocket. Therefore, the STM spectra, combined with the ARPES results, suggest that the pseudogap formation is also orbital dependent. 

It still, however, cannot fully explain the reduction of the gap magnitudes found experimentally and further mechanisms may be needed such as influence of magnetic fluctuations which would further reduce the superfluid stiffness~\cite{Simard2019}, a Lifishitz transition near the orthorhombic to tetragonal transition~\cite{Rhodes2020}, or nematic fluctuations. In FeSe$_{1-x}$S$_x$, the nematic fluctuations are peaked at $q_{\rm nem} \approx 0$ because its ordered state breaks the rotational symmetry while preserving translational symmetry. Enhanced nematic fluctuations~\cite{hosoi16} play an important role for the normal state properties~\cite{Licciardello19,reiss20}, particularly at $x\sim x_c$~\cite{Licciardello19}. It has also been pointed out theoretically that nematic fluctuations strongly influence the superconductivity~\cite{maier14,lederer17}. Therefore, it is tempting to consider that the nematic fluctuations and incipient bands in FeSe$_{1-x}$S$_x$ upon approaching a tetragonal phase can enhance the pairing interaction, leading to the system to approach the BEC regime. Further theoretical and experimental studies are required to uncover whether the nematic fluctuations and multi-band nature affect the crossover physics. 

Our thermodynamic studies have revealed a highly non-mean-field behavior of superconducting transition in FeSe$_{1-x}$S$_x$ when the nematic order of FeSe is completely suppressed by S substitution. This is consistent with the recent ARPES measurements which report the BEC-like dispersion of the Bogoliubov quasiparticles in the tetragonal regime. Our findings demonstrate that FeSe$_{1-x}$S$_x$ offers a unique playground to study the superconducting properties in the BCS-BEC crossover regime of multiband systems. 

\section{ACKNOWLEDGMENTS}

The authors thank A.~Carrington, L.~Malone, P.~Walmsley, K.~Ishida, K.~Sugii, T.~Taen, and T.~Osada for experimental assistance. We also thank Y.~Tada, H.~Ikeda, K.~Adachi, Y.~Ohashi, and D.~Inotani for valuable discussion. This work has been supported by KAKENHI (Nos.~JP20H02600, JP20K21139, JP19H00649, JP18H05227, JP18K13492) and Grant-in-Aid for Scientific Research on Innovative Areas ``Quantum Liquid Crystals'' (No.\ JP19H05824) from JSPS.

\section{APPENDIX A: MATERIALS AND METHODS}
\bigskip
\noindent
{\bf Crystal growth and characterization}
The single crystals of FeSe$_{1-x}$S$_x$ were grown by the chemical vapor transport technique as described in Refs.\,\cite{boehmer13,hosoi16}. The composition $x$ is determined by the energy dispersive X-ray spectroscopy.

\bigskip
\noindent
{\bf Heat capacity}
The heat capacity of the crystals is measured using long relaxation method~\cite{wang01,taylor07}, whose experimental setup is illustrated in the inset of Fig.\:1(a). A single bare chip of Cernox resistor is used as the thermometer, heater and sample stage, which is suspended from the cold stage by silver-coated glass fibers in order that the bare chip has weak thermal link to the cold stage, and electrical connection for the sensor reading. The mass of the samples measured in this study is 81, 41, 171, 137, 179, 50, 107, 192, and 10~$\mu$g for $x$ = 0, 0.05, 0.1, 0.13, 0.16, 0.18, 0.20\#1, 0.20\#2, and 0.21, respectively. The samples are mounted on the bare chip using Apiezon N grease. The heat capacity of the crystals are typically obtained by subtracting the heat capacity of bare chip and grease from the raw data.

\bigskip
\noindent
{\bf Torque}
The torque measurements are performed using micro-cantilevers. The small single crystals with typical size of 150~$\mu$m$\times$100~$\mu$m$\times$10~$\mu$m are mounted on the tip of micro cantilevers with small amount of Apiezon N grease. The magnetic field is applied inside the $ac$($bc$) plane by vector magnet, and the field-angle dependence of torque is obtained by rotating the whole cryostat within $ac$($bc$) plane. Magnetic torque $\tau$ can be expressed as $\tau$ = $\mu_0V\bm{M}\times\bm{H}$, where $\mu_0$ is vacuum permeability, $V$ is sample volume, $\bm{M}$ is magnetization, and $\bm{H}$ is external magnetic field. Here the $\bm{H}$ is rotated within the $ac$ ($bc$) plane, and $\theta$ is polar angle from the $c$ axis. In this configuration, the $\tau$ is expressed as $\tau_{2\theta}(T, H, \theta)$ = $\frac{1}{2}\mu_0H^2V\Delta\chi$sin($\theta$). Here $\Delta\chi$ = $\chi_c$ - $\chi_{ab}$ is anisotropy of magnetic susceptibility between $ac$($ab$) plane and $c$ axis.

\bigskip
\noindent
{\bf Scanning tunneling microscopy/spectroscopy}
The scanning tunneling microscopy/spectroscopy measurements have been performed with a commercial low temperature STM system. Fresh and atomically flat surfaces are obtained by $in$-$situ$ cleavage at liquid nitrogen temperature in ultra high vacuum. All conductance spectra and topographic images are obtained by a PtIr tip. Conductance spectra are measured by a lock-in technique with a modulation frequency 997~Hz and modulation voltage 200~$\mu$V. The backgrounds of raw conductance data are normalized by curves which are derived by 2nd order polynomial fitting using outside superconducting gaps.

\bigskip
\noindent
{\bf Theoretical calculation}
The theoretical calculations were done within a simplified interacting two-band model in two dimensions, employed previously~\cite{chubukov16} with hole and electron pockets with small Fermi energies [Fig.\:8]. We assume superconductivity due to repulsive interactions. While the dominant interband interaction favors Cooper-pairing in the $s$-wave symmetry channel, the $d$-wave projected intraband interaction yields attraction in the d-wave channel. We obtained $T_{pair}$ from the solution of the linearized mean-field gap equations, whereas T$_c$ in each case is estimated from superfluid stiffness. Details of the calculations are given in the Supplemental Information.

\section{APPENDIX B: ENERGY DISPERSIVE X-RAY SPECTROSCOPY (EDX) ANALYSIS}

\begin{figure*}[t]
\centering
\includegraphics[width=0.9\linewidth]{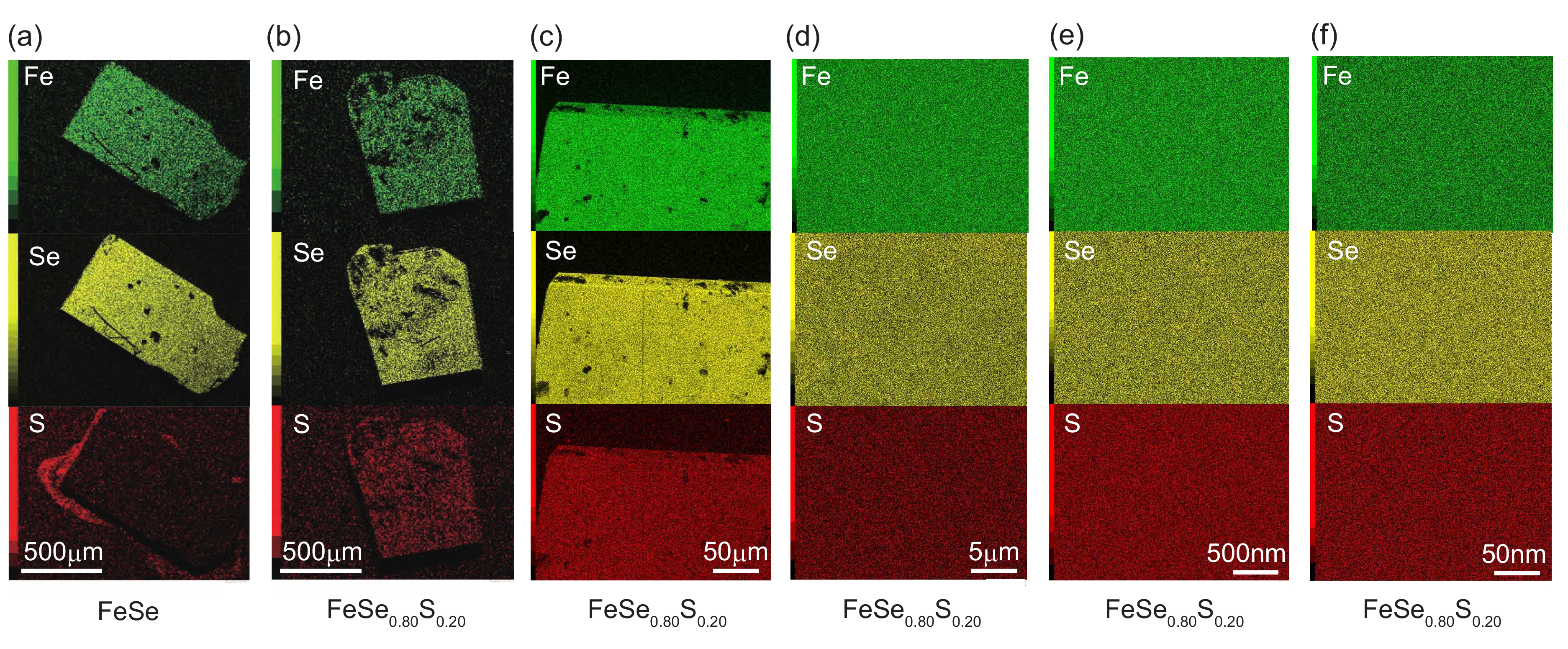}
\vspace{-3mm}
\caption{Energy Dispersive X-ray spectroscopy (EDX) analysis on tetragonal FeSe$_{0.80}$S$_{0.20}$ samples in comparison with FeSe.
	(a) Elemental mapping of Fe (green dots), Se (yellow dots), and S (red dots) at 500~$\mu$m scale for $x$ = 0. (b)-(f) Elemental mapping of Fe, Se, and S for $x$ = 0.20 at the scale of 500~$\mu$m((b)), 50~$\mu$m((c)), 5~$\mu$m((d)), 500~nm((e)), and 50~nm((f)), respectively.
}
\label{figS1}
\end{figure*}

Energy Dispersive X-ray spectroscopy (EDX) analysis is conducted on tetragonal $x$ = 0.20 samples in comparison with $x$ = 0 to investigate the spatial homogeneity of the sulfur content. For $x$ = 0, sulfur intensity is almost same as the background outside the sample position in Fig.\:\ref{figS1}(a), showing that there is no sulfur substituted into the sample as expected. Figures\:\ref{figS1}(b)-(f) show the elemental distribution of $x$ = 0.20 at the scale of 500~$\mu$m,  50~$\mu$m, 5~$\mu$m, 500~nm, and 50~nm, respectively. Each data shows S intensity with the much thicker color than $x$ = 0 although there is some small darker areas in Fig.\:\ref{figS1}(b), and (c) which originate from the large impurity attached on the crystal surface and surface roughness in a macroscopic scale and are not from the inhomogeneity of chemical composition. The data at every length scale show the uniform distribution of sulfur which is almost comparable to iron and selenium, indicating that there is no discernible segregation and inhomogeneity down to the mesoscopic scale $\simeq$10~nm. We quantitatively investigate the sulfur content $x$ at the several area inside the sample and the typical variation of $x$ is $\simeq$0.01, demonstrating that such variation of composition cannot explain the results of specific heat and torque measurements.\\

\section{APPENDIX C: COMPARISON WITH PREVIOUS HEAT CAPACITY DATA}

\begin{figure}[b]
\centering
\includegraphics[width=0.9\linewidth]{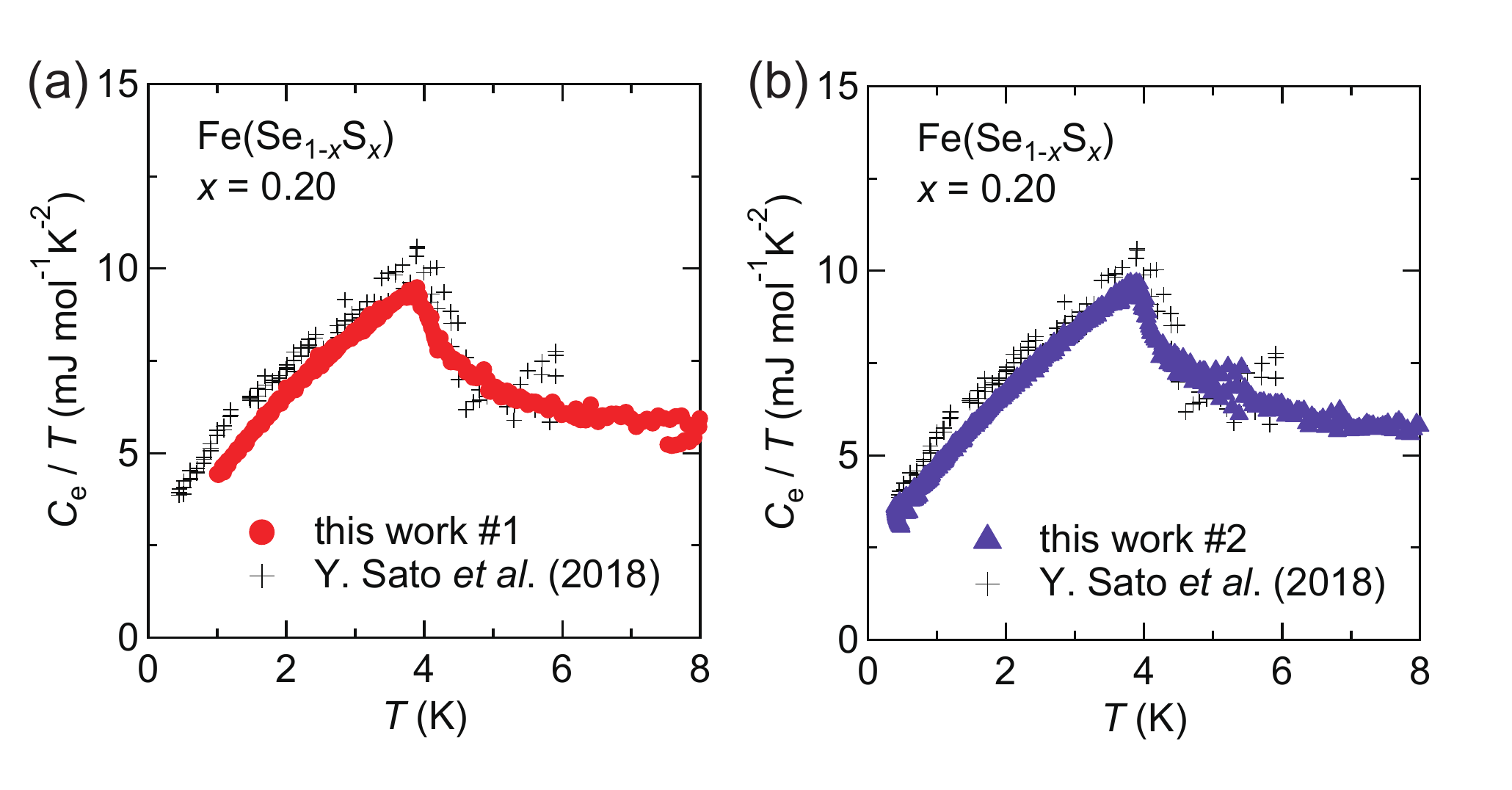}
\vspace{-3mm}
\caption{Temperature dependence of heat capacity in tetragonal FeSe$_{1-x}$S$_x$ samples.
	(a), (b) Temperature dependence of electronic heat capacity divided by temperature for $x$ = 0.20 \#1 ((a)) and \#2 ((b)) in comparison with the data in Refs.\,\cite{sato18}. 
}
\label{figS2}
\end{figure}

The heat capacity in the tetragonal FeSe$_{1-x}$S$_x$ has been reported previously by Y.~Sato et al., using quasi-adiabatic method\cite{sato18}. We plot our $C_e$/$T$ data for two $x$ = 0.20 samples in Figs.\:\ref{figS2}(a), (b) together with the data of Ref.\cite{sato18}. The two samples in this work exhibit almost identical temperature dependence and absolute values, indicating the reproducibility of the heat capacity data in $x$ = 0.20 with better resolution in our studies. Although the absolute value of our data is slightly smaller than the data in Ref.\cite{sato18}, our data is within the error range of previous data, and both data are basically consistent up to $\simeq$ 6 K which is the highest temperature in the previous reports.

\section{APPENDIX D: UPPER CRITICAL FIELD $H_{C2}$ AND ESTIMATION OF GAUSSIAN FLUCTUATION TERM}

\begin{figure*}[t]
\centering
\includegraphics[width=0.8\linewidth]{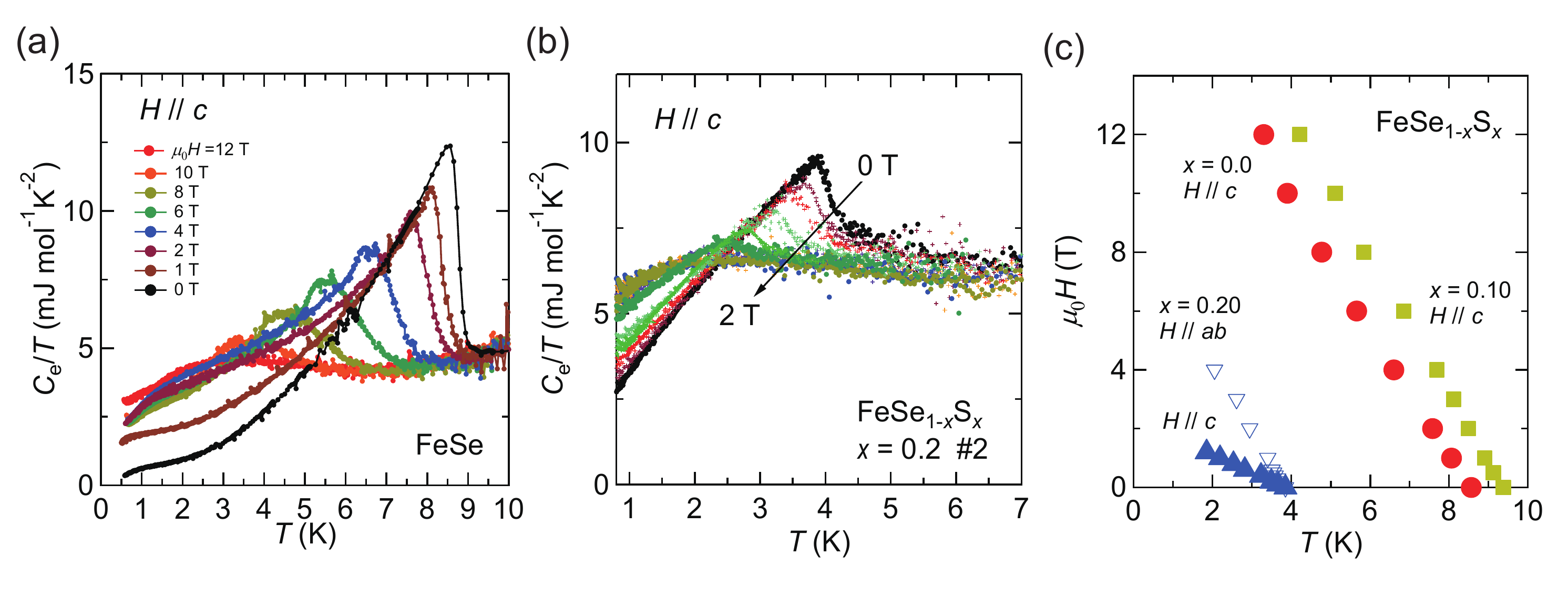}
\vspace{-3mm}
\caption{Heat capacity in magnetic field for FeSe$_{1-x}$S$_x$ $x$ = 0 and 0.20 and $x$-dependence of upper critical field.
	(a), (b) Temperature dependence of electronic heat capacity divided by temperature in magnetic field for $x$ = 0 ((a)) and $x$ = 0.20 \#2 ((b)). The magnetic field is applied along $c$-axis direction for both samples. 
 (c) Temperature dependence of upper critical field $H_{c2}$($T$) determined from the heat capacity in magnetic field. The $H_{c2}$($T$) for $H$ // $ab$ in $x$ = 0.20 and the $H_{c2}$($T$) for $H$ // $c$ in $x$ = 0.10 are also plotted in the same panel.
}
\label{figS3}
\end{figure*}

The field $H$ dependence of heat capacity is measured in the perpendicular field $H // c$ for $x$ = 0.0, and 0.10, and in the perpendicular and parallel field $H // c, ab$ for $x$ = 0.20. The $C_e$/$T$ in $H // c$ for several $H$ values are shown for $x$ = 0.0 and 0.20 in Figs.\:\ref{figS3}(a), (b), respectively. The heat capacity jump due to the superconducting transition becomes broader with increasing $H$ for both S concentrations. This behavior is possibly due to the enhanced fluctuations in the magnetic field as discussed in YBa$_2$Cu$_3$O$_{7-\delta}$\cite{overend94}. In zero field, we observe the large superconducting fluctuations in heat capacity in $x$ = 0.20. The additional heat capacity due to the superconducting fluctuations just above $T_c$ is suppressed by applying magnetic field, and then the $C_e$/$T$ approaches to the normal state value. This fact indicates that the additional heat capacity responds sensitively to the magnetic field, and does not stem from the incorrect estimation of electronic heat capacity in the normal state when we subtract the lattice contribution. From the $T_c$ defined as the peak temperature in $C_e$/$T$ under field, we obtain the $H$-$T$ phase diagram of $x$ = 0.0, 0.10 and 0.20 as shown in Fig.\:\ref{figS3}(c). In $x$ = 0.0 and 0.10 the $T$-dependence of upper critical field $H_{c2}$ is consistent with previous reports\cite{muratov17, abdel15}, while the $H_{c2}$ for $\mu_0H > 12$ T cannot be determined from our measurements. The $H_{c2}$ in $x$ = 0.20 shows small value compared to $x$ = 0.0, reflecting the lower $T_c$. From the $H_{c2}(0)$ estimated by the WHH relation $H_{c2}(0) = 0.69T_c|dH_{c2}/dT|_{T = T_c}$\cite{werthamer66}, we obtain the coherence length $\xi_{ab}=13.5$\,nm, $\xi_c=4.1$\,nm for $x$ = 0.20 through $H_{c2}(0) = \Phi_0/(2\pi\xi_{ab}^2)$, and $\Phi_0/(2\pi\xi_{ab}\xi_c)$ for $H // c$, and $H // ab$, respectively.\\

The contribution of the mean field Gaussian fluctuations to the heat capacity~\cite{inderhees88} is given by $C_{\rm gauss} = C^{+}t^{-0.5}$, where $C^{+} = k_{\rm B}/(8\pi\xi_{ab}^2\xi_c$) and $t\equiv \frac{T-T_{\rm c}}{T_{\rm c}}$ is the reduced temperature, and $\xi_{ab}$ and $\xi_c$ are in-plane and out-of-plane coherence lengths at $T = 0$, respectively. The dashed lines in Figs.\:4(a) and (b) represent the contribution of Gaussian fluctuations obtained by using $\xi_{ab}=5.5$\,nm, $\xi_c=1.5$\,nm for $x=0$~\cite{kasahara16} and $\xi_{ab}=13.5$\,nm, $\xi_c=4.1$\,nm for $x=0.20$ (Fig.\:7).

The Gaussian-type (Aslamasoz-Larkin, AL) fluctuation contribution in susceptibility is given by
\begin{equation}
\chi_{\rm AL}\approx \frac{2\pi^2}{3}\frac{k_{\rm B}T_{\rm c}}{\Phi_0^2}\frac{\xi_{ab}^2}{\xi_c}t^{-\frac{1}{2}}
\end{equation}
in the zero-field limit. Here $\Phi_0$ is the flux quantum. In the multiband case, the behavior of $\chi_{\rm AL}$ is determined by the shortest coherence length of the main band, which governs the orbital upper critical field. As the diamagnetic contribution $\chi_{\rm AL}$ is expected to become smaller in magnitude at higher magnetic fields, $|\chi_{\rm AL}|$ yields an upper bound for the conventional Gaussian-type amplitude fluctuations.

\section{APPENDIX E: CALCULATIONS OF PAIRING INSTABILITY AND PAIRING CHANNEL BASED ON THE TWO BAND MODEL}

\begin{figure*}[t]
\centering
\includegraphics[width=0.8\linewidth]{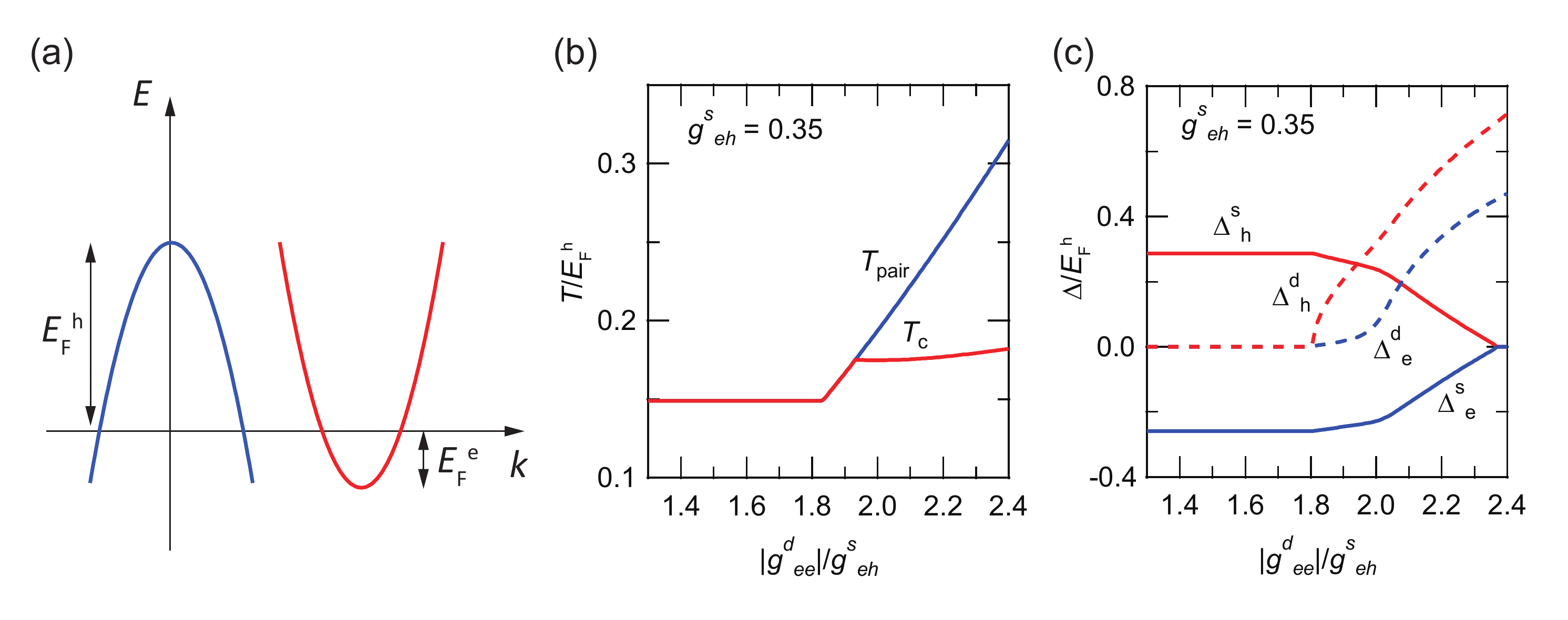}
\vspace{-3mm}
\caption{Calculation results of pairing and condensation temperature, and intensity of pairing channels based on the two band model.
       (a)  Schematic picture of the assumed electronic structure with hole(blue curve) and electron(red curve) bands. The $E^h_{F}$ and $E^e_{F}$ indicate the Fermi energy of the hole and electron band, respectively.
	(b) Calculated pairing temperature $T_{\rm pair}$ and condensation temperature $T_{\rm c}$ in units of $E_{F}^h$ as a function of the ratio of intraband to interband interactions in a two-band model. 
 (c) Calculated superconducting gaps in units of $E_{F}^h$ for $s$-wave and $d$-wave channels in the electron (e) and hole (h) bands as a function of the ratio of intraband to interband interactions.
}
\label{figS4}
\end{figure*}

To make an estimate of the splitting between the Cooper-pair formation temperature, $T_{pair}$, and the actual superconducting transition temperature, $T_c$, we followed Ref.\cite{chubukov16} and consider a simplified interacting two-band model in two dimensions with hole and electron pockets with small Fermi energies ($E_{F}^h =20$meV and $E_{F}^e =10$meV) as depicted in the Fig.\:\ref{figS4}(a). As we are interested in the evolution of the FeSe system as a function of the  Sulfur doping we assumed the system is orthorhombic. However, once the intraband interaction in the $d$-wave interaction dominates and the $s-$wave component of the gap is small the results can be easily extrapolated to the purely tetragonal system. The Hamiltonian reads
\begin{eqnarray*}
\begin{split}
\hat{H} = & \sum_{\kk\alpha\sigma}\xi_{\kk\alpha}c^\dagger_{\alpha\kk\sigma}c_{\alpha\kk\sigma} \\
 &+ \sum_{\kk,\kk^\prime,\alpha,\alpha^\prime}U_{\alpha\alpha^\prime}(\kk\kk^\prime)c^\dagger_{\alpha\kk\ua}c^\dagger_{\alpha-\kk\da}c_{\alpha^\prime-\kk^\prime\da}c_{\alpha^\prime\kk^\prime\ua}
\end{split}
\end{eqnarray*}
where $\alpha\in\{\mathrm{e},\mathrm{h}\}$, and $\xi_{\kk \mathrm{e}}$, $\xi_{\kk \mathrm{h}}$ are the electron and hole energy dispersions separated by the large momentum, respectively as shown in the Fig.\:\ref{figS4}(a).
Assuming superconductivity due to repulsive interaction in the $A_{1g}$ and $B_{2g}$ symmetry channels we write the interaction terms as follows
\begin{eqnarray}
U_{\mathrm{eh}}(\kk,\kk^\prime) & = & U^s_{\mathrm{eh}}+U^d_{\mathrm{eh}}\phi(\kk)\phi(\kk^\prime) \nonumber \\
U_{\alpha\alpha}(\kk,\kk^\prime) &  = & U^d_{\alpha\alpha}\phi(\kk)\phi(\kk^\prime) \nonumber
\end{eqnarray} 
with $\phi(\kk)=\cos(2\varphi)=\frac{k_x^2-k_y^2}{k_x^2+k_y^2}$. 
Here, we further assume that the inter-band repulsion drives $s^{\pm}$-wave symmetry of the superconducting order parameter, while interaction in the $d$-wave channel is mostly intra-band. 
We define superconducting order parameters as
\begin{align}
\Delta_{\alpha}(\kk)=\sum_{\kk^\prime,\alpha\prime}U_{\alpha\alpha\prime}(\kk,\kk^\prime)\langle c_{\alpha^\prime-\kk^\prime\da}c_{\alpha^\prime\kk^\prime\ua}\rangle
\end{align}
and perform a mean-field decoupling to find the mean-field gap equations
\begin{eqnarray*}
\begin{split}
\Delta^s_{\mathrm{e}}=&-g^s_{\mathrm{eh}}\int_{0}^{2\pi}\frac{d\varphi}{2\pi}\int_0^{\Lambda}d\epsilon\frac{\tanh\left(\frac{E_{\mathrm{e}}(\kk)}{2T}\right)}{2E_{\mathrm{e}}(\kk)}\nonumber\\
\Delta^s_{\mathrm{h}}=&-g^s_{\mathrm{eh}}\int_{0}^{2\pi}\frac{d\varphi}{2\pi}\int_0^{\Lambda}d\epsilon\frac{\tanh\left(\frac{E_{\mathrm{h}}(\kk)}{2T}\right)}{2E_{\mathrm{h}}(\kk)}\nonumber\\ 
\Delta^d_{\mathrm{e}}=&|g^d_{\mathrm{ee}}|\int_{0}^{2\pi}\frac{d\varphi}{2\pi}\int_0^{\Lambda}d\epsilon\frac{\cos^2(\varphi)\tanh\left(\frac{E_{\mathrm{e}}(\kk)}{2T}\right)}{2E_{\mathrm{e}}(\kk)} \\
& -g^d_{\mathrm{eh}}\int_{0}^{2\pi}\frac{d\varphi}{2\pi}\int_0^{\Lambda}d\epsilon\frac{\cos^2(\varphi)\tanh\left(\frac{E_{\mathrm{h}}(\kk)}{2T}\right)}{2E_{\mathrm{h}}(\kk)}\nonumber\\
\Delta^d_{\mathrm{h}}=&|g^d_{\mathrm{hh}}|\int_{0}^{2\pi}\frac{d\varphi}{2\pi}\int_0^{\Lambda}d\epsilon\frac{\cos^2(\varphi)\tanh\left(\frac{E_{\mathrm{h}}(\kk)}{2T}\right)}{2E_{\mathrm{h}}(\kk)} \\
&-g^d_{\mathrm{eh}}\int_{0}^{2\pi}\frac{d\varphi}{2\pi}\int_0^{\Lambda}d\epsilon\frac{\cos^2(\varphi)\tanh\left(\frac{E_{\mathrm{e}}(\kk)}{2T}\right)}{2E_{\mathrm{e}}(\kk)}\nonumber.\\\label{GapEuqation}
\end{split}
\end{eqnarray*}
The dimensionless coupling constants are now given by dimensionless $g_{\alpha\alpha\prime}=N_0U_{\alpha\alpha\prime}$ with the density of states in two dimensions $N_0=\frac{m}{2\pi}$, $\Lambda = 1$eV $\gg E_{F}^h$ is the high energy cut-off and $E_{\alpha}(\kk)=\sqrt{\xi_{\alpha}^2+\left[\Delta^s_{\alpha}+\Delta^d_\alpha\cos(\varphi)\right]^2}$ are the energy dispersion of the Bogoliubov quasiparticles. 
Note that while the inter-band term between electron and hole pockets is assumed to be repulsive, the intraband $g^d_{\mathrm{ee}}<0$ and $g^d_{\mathrm{hh}}<0$ are attractive in the $d$-wave channels.
In case of $\Delta_{\alpha}/E_{\mathrm{F}\alpha}\sim 1$, where $E_{\mathrm{F}\alpha}$ is the Fermi energy of each band, we need to renormalize the chemical potential assuming the total number of particles is conserved.
The equation determining the particle number is given by
\begin{align}
\begin{split}
E_{\mathrm{F_e}}-E_{\mathrm{F_h}}=&-\int_0^{2\pi}\frac{d\varphi}{2\pi}\int_{0}^{\Lambda}d\epsilon \\
&\left(\frac{\xi_e(\kk)\tanh\left(\frac{E_{\mathrm{e}}(\kk)}{2T}\right)}{2E_{\mathrm{e}}(\kk)}+\frac{\xi_h(\kk)\tanh\left(\frac{E_{\mathrm{h}}(\kk)}{2T}\right)}{2E_{\mathrm{h}}(\kk)}\right)\label{NumberEq}
\end{split}
\end{align}
and it has to be solved self-consistently with Eqs.(\ref{GapEuqation}).
The pair building temperature $T_{\mathrm{pair}}$ at which electrons form Cooper pairs can be obtained from the condition that the determinant in
\begin{align}
0=-\left(\begin{array}{cccc}
1& g^s_{\mathrm{eh}}\Pi^s_{h} &  &  \\ 
g^s_{\mathrm{eh}}\Pi^s_{e} &1  &  &  \\ 
&  & 1-|g^d_{\mathrm{ee}}|\Pi^s_{e} & g^d_{\mathrm{eh}}\Pi^d_{e} \\ 
&  & g^d_{\mathrm{eh}}\Pi^d_{e} & 1-|g^d_{\mathrm{hh}}|\Pi^s_{h}
\end{array} \right)
\left(\begin{array}{c}
\Delta^s_{\mathrm{e}} \\ 
\Delta^s_{\mathrm{h}} \\ 
\Delta^d_{\mathrm{e}} \\ 
\Delta^d_{\mathrm{h}}
\end{array} \right)\label{LinearisedGapEquations}
\end{align}
vanishes and
\begin{align}
\Pi^s_{\alpha} &= \int_{0}^{2\pi}\frac{d\varphi}{2\pi}\int_{0}^{\Lambda}d\epsilon\frac{\tanh\left(\frac{\xi_\alpha}{2T_{\mathrm{ins}}}\right)}{2\xi_\alpha}\\
\Pi^d_{\alpha} &= \int_{0}^{2\pi}\frac{d\varphi}{2\pi}\int_{0}^{\Lambda}d\epsilon\frac{\cos^2(2\varphi)\tanh\left(\frac{\xi_\alpha}{2T_{\mathrm{pair}}}\right)}{2\xi_\alpha}.
\end{align}	
In the usual BCS case the phase fluctuations are costly	and $T_{\mathrm{c}}\approx T_{\mathrm{pair}}$. In our case due to smallness of the Fermi energies, the condensation of pairs may happen at lower temperature, $T_{\mathrm{c}}\leq T_{\mathrm{pair}}$. We estimate $T_{\mathrm{c}}\approx\frac{\pi}{2}\rho_{\mathrm{s}}$ where $\rho_{\mathrm{s}}(T=0)$ is the superfluid-stiffness following Ref.~\cite{chubukov16}. Note that in two dimensions, the superconducting
transition temperature $T_c \sim \rho_s(T_c)$ (see, e.g., Refs.\cite{Pokrovsky79,Beasley79}).
The interplay between $T_c$ and $T_{pair}$ depends on the ratio
$\rho_s (T = 0)/T_{pair}$. If this ratio is large, the superfluid stiffness rapidly increases below T$_{pair}$. In this situation, $T_c = T_{pair}$ minus a small correction, i.e., the phases of bound pairs order almost
immediately after the pairs develop (phase fluctuations cost
too much energy). If $\rho_s (T = 0)/T_{pair}$ is small, $\rho_s (T)$ increases slowly below T$_{pair}$ and $T_c$ is of order $\rho_s (T = 0)$. In Ref.\cite{chubukov16} its expression was extended to the multiband case where $\rho_{\mathrm{s}}$ is given by $\rho_{\mathrm{s}}\approx \rho_{\mathrm{e}}+\rho_{\mathrm{h}}$,
\begin{align}
\rho_{\mathrm{e}}&=\frac{1}{16\pi^2}\int_0^{2\pi}d\varphi\int_{0}^{\Lambda}d\epsilon\Delta_{\mathrm{e}}^2\frac{\epsilon}{E_{\mathrm{e}}(\kk)^3}\label{RhoE}\\
\rho_{\mathrm{h}}&=\frac{1}{16\pi^2}\int_0^{2\pi}d\varphi\int_{0}^{\Lambda}d\epsilon\Delta_{\mathrm{h}}^2\frac{\epsilon}{E_{\mathrm{h}}(\kk)^3}.\label{RhoH}
\end{align}

We solved Eqs.~(\ref{GapEuqation}) together with Eq.~(\ref{NumberEq}) for the mixed $s+d$-wave gaps in the orthorhombic state  at $T=0$ and used them as an input parameter to calculate $T_{\mathrm{c}}$.
The Cooper-pair formation temperature, $T_{\mathrm{pair}}$ is found from the condition that the  determinant in Eq.~(\ref{LinearisedGapEquations}) vanishes and the results are shown in the Fig.\:\ref{figS4}(b) in units of $E_{Fh}$. The pair-formation temperature $T_{\rm pair}$ and condensation temperature $T_{\rm c}$ are calculated as a function of the ratio of intraband and interband interactions $|g^d_{\rm ee}|/g^s_{\rm eh}$ shown in Fig.\:\ref{figS4}(b). The two temperatures split and the difference grows with increasing $|g^d_{\rm ee}|/g^s_{\rm eh}$. Fig.\:\ref{figS4}(c) shows the evolution of the gap symmetry as a function of $|g^d_{\rm ee}|/g^s_{\rm eh}$ also in units of $E_{Fh}$, and we can see the dominant $d$-wave character for each band with large $|g^d_{\rm ee}|/g^s_{\rm eh}$ regime. Observe that for smaller ratio of $|g^d_{ee}|/g^s_{eh}$ the pairing symmetry is a pure $s^{+-}$ driven by inter-band interactions. In this case both gaps are equal in magnitude, but with the larger gap at the band with smaller $E_F$. With increasing intra-band coupling in the d-wave gaps $\Delta^d_e$ and  $\Delta^d_h$ grow while the s-wave gaps become smaller. Since we assume $|g^d_{ee}|\gg g^d_{eh}$, $\Delta^d_e$ and $\Delta^d_h$ differ in magnitude with the larger gap at the band with larger $E_F$.



\end{document}